\documentclass[twocolumn,nofootinbib,aps,pra,10pt]{revtex4-1}
\usepackage{amssymb}
\usepackage{graphicx}
\usepackage{amsmath}
\usepackage{color}
\usepackage{subfigure}
\usepackage{mathrsfs}
\usepackage{bm}

\newcommand{\rr}{\mathbf{r}}

\newcommand{\dd}{\text{d}}

\newcommand{\OM}{\mathbf{\Omega}}
\newcommand{\pp}{\mathbf{p}}
\newcommand{\unit}[1]{\mathbf{\hat{#1}}}
\newcommand{\diff}[2]{\frac{\dd #1}{\dd #2}}
\newcommand{\pdiff}[2]{\frac{\partial #1}{\partial #2}}

\usepackage{times}
\usepackage{amsmath}
\usepackage{amsfonts}
\usepackage{amssymb}
\usepackage{bm}
\usepackage{graphicx}
\usepackage{epstopdf}
\usepackage{enumerate}
\usepackage{color}

\usepackage{braket}

\graphicspath{{img/}}

\begin{document}
\title{Quasi-Classical Rules for Qubit Spin-Rotation Error Suppression.}
\author{Qile David Su}
\affiliation{$^{1}$Department of Physics and Astronomy, University of California, Los Angeles, CA 90095, USA}

\begin{abstract}
A frequently encountered source of systematic error in quantum computations is imperfections in the control pulses which are the classical fields that control qubit gate operations. From an analysis of the quantum-mechanical time-evolution operator of the spin wavefunction, it has been demonstrated that \textit{composite} pulses can mitigate certain systematic errors and an appealing geometric interpretation was developed for the design of error-suppressing composite pulses. Here we show that these same pulse sequences can be obtained within a quasi-classical framework. This raises the question of whether error-correction procedures exist that exploit entanglement in a manner that can not be reproduced in the quasi-classical formulation. 
\end{abstract}
\maketitle

\section{Introduction}
An elementary single-qubit quantum gate is the X gate, which is one possible quantum analog of a classical NOT gate. This is implemented by applying radiation resonant with the qubit transition with a pulse area of $\pi$, commonly known as a $\pi$-pulse. During such a pulse, the Bloch vector, a geometrical representation of the state of the qubit, rotates by an angle of $\pi$. In practice, imperfect control of pulse amplitude, duration, frequency, and phase leads to gate infidelity \cite{low_optimal_2014}. Coupling between a qubit and its environment generates inelastic interactions that can introduce random errors through phase decoherence \cite{steane_error_1996, shor_scheme_1995, manzano_short_2020}. This paper will be restricted to two frequently encountered systematic errors namely imperfect control of the amplitude of the pulse, or \textit{amplitude error}, and imperfect control of the frequency of the pulse, or \textit{detuning error} \footnote{Depending on the context, amplitude errors may also be known as flip-angle errors or pulse-length errors, while detuning errors are also known as off-resonance errors or frequency offsets.}. 

Work by the NMR community showed that amplitude and detuning errors can be reduced by replacing a single $\pi$-pulse by \textit{composite} pulses if the errors are assumed to be constant throughout the pulse \cite{ryan_robust_2010, souza_robust_2011}. It was shown \cite{low_optimal_2014} that an amplitude error of order $\epsilon$ incurred during the rotation of a Bloch vector by $\pi$ can -- at least in principle -- be reduced to an error of the order of $\epsilon^n$, with $n$ an integer, by breaking the $\pi$-pulse up into a sequence of $2n$ rotations of the Bloch vector over $\pi$ or $2\pi$. Subsequently, general constraints were derived on the control pulses that, if obeyed, should reduce the amplitude and detuning error, while a geometrical interpretation of these constraints was developed as well \cite{jones_designing_2013, merrill_transformed_2014}. Mathematically, the method relies on the \textit{Magnus expansion} \cite{blanes_magnus_2009}, which provides an exponential representation of the time-evolution operator (or propagator) of the spin wavefunction. Similar methods have been developed to study the effects of disorder and spin-spin interaction \cite{choi_robust_2020}. An experimental search for sequences that simultaneously suppress the amplitude and detuning error \cite{souza_robust_2011} produced the so-called \textit{Knill sequence}, a seemingly complex sequence of five $\pi$-pulses discussed below. Using the method of refs.\cite{jones_designing_2013, merrill_transformed_2014}, it can be shown that the Knill sequence is a member of a broader, one-parameter family of sequences that eliminate the leading order amplitude and detuning errors simultaneously. 

The spin dynamics underlying NMR (as well as MRI and ESR) is traditionally described by the \textit{Bloch equations}. While the derivation of these equations is based on quantum mechanics, they have the form of the equation of motion of a (quasi-) classical magnetic moment precessing in a magnetic field (coupling to the environment produces damping terms). This suggests that it should be possible to frame the error correction method of ref.\cite{jones_designing_2013, merrill_transformed_2014} in the quasi-classical language of the precession of a magnetic moment instead of the propagator method. Such a construction could provide a better intuitive insight into the -- rather complex -- multi-step error-suppressing sequences. In this paper we develop such a quasi-classical framework for the case of composite sequences composed of $\pi$-pulses that correct for amplitude and detuning errors. Through a straightforward perturbative analysis, we obtain constraints on a pulse sequence that, if obeyed, suppress amplitude and detuning errors to second order for the expectation value of the spin. Next we show that the resulting pulse sequences are equivalent to those derived from the framework of ref.\cite{jones_designing_2013, merrill_transformed_2014}. The construction of error-suppressing pulse sequences using the quasi-classical framework is sufficiently simple and graphical that it can be part of an upper-division undergraduate class.  

Section II develops a second-order perturbation expansion for the equation of motion of the expectation value of the spin angular momentum due to perturbations generated by amplitude and detuning errors. Expressions are derived for the two global constraints that must be obeyed if first- and second-order amplitude and detuning errors are to be eliminated, respectively. In Section III the first-order constraint is illustrated for a variety of sequences and the equivalence between the propagator and quasi-classical frameworks is demonstrated. In Section IV, the second-order constraint is discussed similarly. In Section V, we apply the quasi-classical framework to more general rotations. In Section VI we provide the formal connection between the quasi-classical perturbation expansion.

\section{Dynamics of the Bloch Vector and Perturbation Theory.}
\label{sec.dynamics}
To compute how errors in the control pulse accumulate during a pulse sequence, we will follow the evolution of a single qubit. The specific realization of a qubit considered in this paper is a spin-1/2 two-level system subject to a DC magnetic field along the $z$ direction. The level splitting is $\hbar\omega_0$ with $\omega_0$ the Larmor frequency. Let the spinor $\ket{\psi}=(\alpha,\beta)$ denote the general spin state $\alpha\ket{\uparrow}+\beta\ket{\downarrow}$. A $\pi$-pulse is performed by the application of an AC magnetic field rotating in the $xy$-plane with frequency $\omega$ and phase $\phi$ \cite{vandersypen_nmr_2005}. In terms of the Pauli matrices, the spin Hamiltonian can be expressed as
\begin{equation}
    H=\hbar/2[\omega_0\hat\sigma_z+\Omega_0\cos(\omega t+\phi)\hat\sigma_x+\Omega_0\sin(\omega t+\phi)\hat\sigma_y],
\end{equation}
where $\Omega_0$ is the magnitude of the AC magnetic field in frequency units. A perfect $\pi$-pulse has frequency $\omega=\omega_0$ and duration $\pi/\Omega_0$. If the drive frequency $\omega$ differs from $\omega_0$, then a detuning error results. If the pulse duration differs from $\pi/\Omega_0$, or if the actual AC field strength differs from $\Omega_0$, an amplitude error results. After transforming into the rotating frame (see appendix \ref{sec.setup-math} for details), the effective Hamiltonian reduces to that of a spin-1/2 particle in a time-independent magnetic field:
\begin{equation}
    H'=\hbar/2[\Delta\ \hat\sigma_z+\Omega_0\cos\phi\ \hat\sigma_x+\Omega_0\sin\phi\ \hat\sigma_y].
\end{equation}
This field $\OM=(\Delta,\Omega_0\cos\phi,\Omega_0\sin\phi)$ (in frequency units) has a $z$ component $\Delta=\omega_0-\omega$ that is the detuning error while the projection of the field in the $xy$-plane makes an angle $\phi$ with the $x$-axis. For convenience, this reference frame will still be referred to as the ``lab frame" even though it rotates with respect to the real laboratory frame. In the following, we will use units with dimensionless time (i.e.,$\ \Omega_0 t$) and dimensionless transverse field strength. 

In appendix \ref{sec.setup-math} it is shown that if the Ehrenfest Theorem is applied to the expectation value $\rr\equiv {\braket{\boldsymbol\sigma}}$ of the spin operator -- i.e., the Bloch vector -- in a magnetic field then this produces an equation of motion that is the same as that for the precession of a classical magnetic moment in a magnetic field: 
\begin{equation}
    \diff{\rr(t)}{t}=\OM(t)\times\rr(t).
    \label{nominal-eq}
\end{equation}
Note that Planck's constant no longer appears. This equation is exact in the absence of coupling to the environment. Moreover, it is similar to the classical mechanics of rigid-body rotation and easy to visualize. We will use it as a starting point for the construction of pulse sequences that compensate for detuning and/or amplitude errors. To that purpose we first separate the error-free (or ``nominal") part of the pulse $\OM(t)$ from the amplitude and detuning errors $\OM_1(t)$, so
\begin{equation}
    \diff{\rr(t)}{t}=(\OM(t)+\OM_1(t))\times\rr(t).\label{master-eq}
\end{equation}
Amplitude errors are represented by the time-dependent error vector $\OM_1(t)=\epsilon\OM(t)$. They change the rotation rate induced by the control pulse by a factor of $(1+\epsilon)$. Detuning errors are represented as $\OM_1(t)=\Delta\ \unit{z}$, so they are constant and always point along the $z$ direction. The quantities $\epsilon$ and $\Delta$ will be assumed to be constant throughout the sequence and to be small compared to one. 

\begin{figure}[t]
	\begin{minipage}{.45\textwidth}
    	\includegraphics[scale=1]{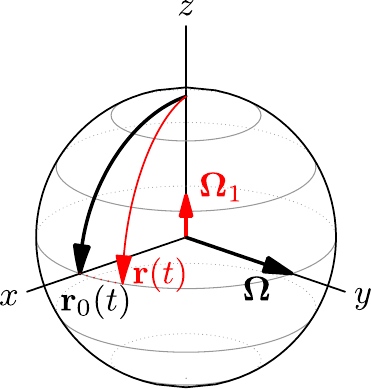}
        \caption{Motion of the Bloch vector in the lab frame for a $\pi/2$ pulse about the $y$-axis in the presence of a detuning error [see Eq.\eqref{master-eq}]. Black trajectory: nominal motion $\rr_0(t)$ for zero detuning error. Red trajectory: motion in the presence of a detuning error $\OM_1$. The grey circles on the Bloch sphere are contours of constant angle from the spin up direction.}
        \label{fig:before-transform}
    \end{minipage}%
    \hfill
    \begin{minipage}{.45\textwidth}
    	\includegraphics[scale=1]{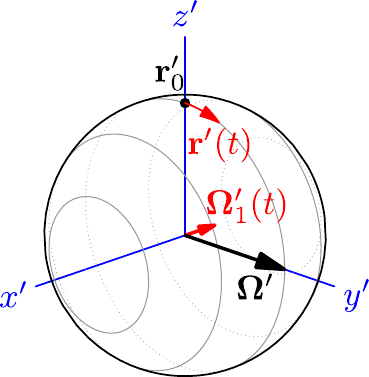}
        \caption{Motion of the Bloch vector (Eq.\eqref{transformed-eq}) in the toggling frame (blue, primed axes) for  the same $\pi/2$ rotation about the lab frame $y$-axis as in figure \ref{fig:before-transform}. Red vector: the detuning error vector $\OM_1'(t)=\Delta \ \mathbf{\hat R}^{-1}(t)\mathbf{\hat{z}}$, which is time-varying. Red trajectory: motion of the Bloch vector $\rr'(t)$ in the toggling frame. The grey circles show the orientation of the Bloch sphere in the toggling frame for the final state, corresponding to the grey circles in the previous figure.}
        \label{fig:after-transform}
    \end{minipage}
\end{figure}

How does the presence of $\mathbf{\Omega}_1(t)$ in the equation of motion modify the evolution of the Bloch vector? Figure \ref{fig:before-transform} shows the motion of the Bloch vector for the case of a $\pi/2$ pulse with a transverse field along the $y$ direction. The initial orientation of the spin is again along the $z$ direction. The black trajectory is the ``nominal" (i.e., error-free) case while the red trajectory shows the deviation induced by a non-zero detuning error. The time-evolution of the nominal motion complicates the visualization of the induced error, and it is a problem that becomes only more serious as the complexity of the sequence increases. This can be circumvented by introducing what is known as the \textit{toggling frame}. This is an ``intermediate" representation that subtracts out the nominal motion caused by $\OM(t)$. Denote the time-dependent transformation matrix between the lab frame and the toggling frame as ${\mathbf{\hat R}}(t)$ so a vector $\mathbf{v}$ in the lab frame (unprimed) has components $\mathbf{v}'(t)$ in the toggling frame (primed). Then we must have that $\mathbf{v}(t)=\mathbf{\hat R}(t)\mathbf{v}'(t)$. In particular, for the magnetic moment $\rr'(t) = \hat{\mathbf{R}}^{-1}\rr(t)$, $\OM'(t) = \hat{\mathbf{R}}^{-1}\OM(t)$, and for the error vector $\OM'_1(t) = \hat{\mathbf{R}}^{-1}\OM_1(t)$. At $t=0$, we choose $\mathbf{\hat R}(0)=\mathbf{I}$, so that $\rr'(0)=\rr(0)$. 

In appendix \ref{sec.toggling-frame} it is shown that the equation of motion of the Bloch vector in the toggling frame is
\begin{equation}
    \diff{\rr'(t)}{t}=\OM_1'(t)\times\rr'(t).
    \label{transformed-eq}
\end{equation}
The amplitude error $\OM_1(t)$, which is time-dependent in the lab frame, becomes piecewise-constant after transformation into $\OM_1'(t)$ in the toggling frame. Conversely, while the detuning error is time-independent in the lab frame, it becomes time-dependent in the toggling frame. Figure \ref{fig:after-transform} compares the motion of the Bloch vector in the lab frame for the case of detuning error with the motion in the toggling frame.

The next step is to solve for the motion of the Bloch vector $\rr'(t)$ in the toggling frame. According to Eq.\eqref{transformed-eq}, if $\OM_1'(t)$ is small then $\rr'(t)$ is nearly constant. This motivates us to construct a perturbation expansion for $\rr'(t)$ in the form, 
\begin{equation}
    \rr'(t)=\rr_0'(t)+\rr_1'(t)+\rr_2'(t)+...
\end{equation}
where terms with subscript $n$ are proportional to the $n$th power of the dimensionless parameter $||\OM_1'||_{\mathrm{max}}t$ (see appendix \ref{sec.convergence} for a discussion of the convergence of the series). Inserting the expansion into Eq.\eqref{transformed-eq} gives
\begin{multline}
    \diff{\rr'_0(t)}{t}+\diff{\rr'_1(t)}{t}+\diff{\rr'_2(t)}{t}+\cdots= \\ 
    \OM_1'(t)\times(\rr'_0(t)+\rr'_1(t)+\rr'_2(t)+\cdots).
\end{multline}
Equating terms of the same order on the two sides of the equation produces a set of recursive equations for successive terms of the perturbation expansion,
\begin{align}
    \diff{\rr_0'(t)}{t}&=0\\
    \diff{\rr_1'(t)}{t}&=\OM_1'(t)\times\rr_0'(t)\\
    \diff{\rr_2'(t)}{t}&=\OM_1'(t)\times\rr_1'(t)\\
    &...\nonumber
\end{align}
Integrating the zeroth-, first- and second-order equations gives
\begin{align}
    \rr_0'(t)&=\rr_0(0)\\
    \rr_1'(t)&=\int_0^t\OM_1'(s)\times\rr_0(0) \mathrm{d}s\\
    \rr_2'(t)&=\int_0^t\OM_1'(s)\times\rr_1'(s) \mathrm{d}s\\
    &...\nonumber
\end{align}
with $\rr_0(0)$ equal to the initial orientation of the Bloch vector $\rr(0)$. Note that each term in the series involves the accumulated action of $\OM_1'(t)$ on the previous term in the series over the time interval $[0,t]$. Crucially, $\OM_1'(t)=\hat{\mathbf{R}}^{-1}\OM_1(t)$ depends on the composite pulse through $\hat{\mathbf{R}}$. Thus given the same errors in the control pulses (which fixes $\OM_1(t)$), the evolution of the Bloch vector in the toggling frame can be drastically different from one composite sequence to the other.

Suppose that one has solved for the time-evolution of successive terms of the expansion of Bloch vector in the toggling frame. Then, at the end of a pulse sequence with total duration $t_{\mathrm{f}}$, the Bloch vector is
\begin{equation}
    \rr'|_{t=t_{\mathrm{f}}}=\rr_0(0)+\rr_1'|_{t=t_{\mathrm{f}}}+\rr_2'|_{t=t_{\mathrm{f}}}+...
\end{equation}
We will say that a sequence suppresses error to the $n$th order if $\rr_k'(t_{\mathrm{f}})=0$ for $k=1,2,...n$. Specifically, a sequence suppresses first-order errors if
\begin{align}
    \rr_1'|_{t = t_{\mathrm{f}}} = \int_0^{t_{\mathrm{f}}}\OM_1'(s)\times\rr_0(0) \mathrm{d}s=0\quad \text{$\forall$}\medspace\medspace \rr_0(0)& \label{first-order} \\
    \text{(First-order constraint)}& \nonumber
\end{align}
and to second order if additionally
\begin{align}
   \rr_2'|_{t = t_{\mathrm{f}}} = \int_0^{t_{\mathrm{f}}}\OM_1'(s)\times\rr_1'(s) \mathrm{d}s=0\quad \text{$\forall$}\medspace\medspace \rr_0(0)& \label{second-order} \\
 \text{(Second-order constraint)}& \nonumber
\end{align}
If these conditions are satisfied for all initial states $\rr(0)$ then the pulse sequence is said to provide universal error correction \cite{low_optimal_2014} or be fully-compensating \cite{vandersypen_nmr_2005}.

\section{First-Order Constraint.}
In this section we focus on the first-order constraint, illustrate how it works for several different sequences and then derive general geometrical rules for first-order error suppression. These rules are then compared with the ones obtained in refs.\cite{jones_designing_2013, merrill_transformed_2014}.

First define the ``error integral" $\pp(t)$ to be 
\begin{equation}
    \pp(t)=\int_0^t\OM_1'(s)\dd s,
    \label{pt-definition}
\end{equation}
where the upper bound of the integration is allowed to vary. The function $\pp(t)$ can be viewed as describing the trajectory of a phantom particle moving in the toggling frame with velocity $\OM_1'(t)$, starting from the origin at $t=0$. The first-order constraint Eq.\eqref{first-order} can be expressed as,
\begin{equation}
    \rr_1'|_{t=t_{\mathrm{f}}}=\pp|_{t=t_{\mathrm{f}}}\times\rr_0(0)=0.
    \label{first-order-manip}
\end{equation}
Define separate-step ``error integral vectors" $\pp_i$ for the individual steps of a pulse sequence:
\begin{equation}
	\pp_i=\int_{t_i}^{t_{i+1}}\OM_1'(s)\dd s,
\end{equation}
with $t_i$ and $t_{i+1}$ the initial and final times of pulse $i$. The first-order constraint can then be written as
\begin{equation}
    \rr_1'|_{t=t_{\mathrm{f}}}=\left(\sum_{i}^N\pp_i\right)\times\rr_0(0)=0.
    \label{first-order-final}
\end{equation}
If the trajectory $\pp(t)$ is closed, then the first-order constraint is obeyed for all initial conditions. Suppose a sequence is composed of $N$ steps. The vectors $\pp_i$ of the individual steps define an $N$-step walk when placed head-to-tail. A sequence is fully-compensating to first order if this $N$-step walk is closed.

\subsection{Spin-echo Sequence}
\begin{figure}[t]
	\includegraphics[scale=1]{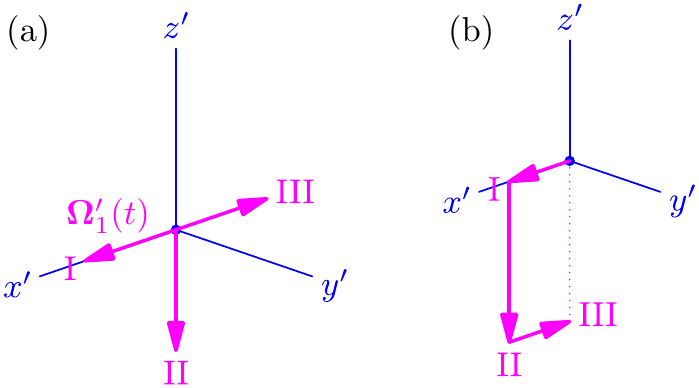}
    \caption{(a): the amplitude error $\OM_1'(t)$ (pink) for the spin-echo sequence $R_0^{\pi/2}\rightarrow R_{\pi/2}^{\pi}\rightarrow R_0^{\pi/2}$ in the toggling frame. The three steps of the sequence are labelled I, II, and III. All three vectors should have length $\epsilon$. (b): the three-step walk consists of the error integral vectors $\pp_i$ for the spin-echo sequence with amplitude error.}
    \label{fig:dOm-pt-amplitude-spinecho}
\end{figure}

Our first example is the three-step \textit{spin-echo} sequence, which is a standard NMR method \cite{levitt_nmr_1979}. The short-hand notation $R_{\phi}^{\theta}$ will be used to denote a single pulse, with $\phi$ the direction of the pulse in the $xy$-plane and $\theta$ the duration of the pulse. An error-free $\pi$-pulse around $x$ is denoted as $R_0^{\pi}$. The spin echo sequence is denoted as $R_0^{\pi/2}\rightarrow R_{\pi/2}^{\pi}\rightarrow R_0^{\pi/2}$ (these angles are in the lab frame). It has the following time-dependent pulse direction $\OM(t)$:
\begin{equation}
    \OM(t)=\begin{cases}
    \unit{x} & t\in(0,\frac{\pi}{2})\\
    \unit{y} & t\in(\frac{\pi}{2},\frac{3\pi}{2})\\
    \unit{x} & t\in(\frac{3\pi}{2},2\pi)\end{cases}.
    \label{spin-echo-spec}
\end{equation}
In the presence of amplitude or detuning error, the first step is calculating the error $\OM_1'(t)$ in the toggling frame. This is easy for the amplitude error case because the error is always parallel to the pulse direction $\OM'(t)$, which in turn is piecewise constant in the toggling frame. The three vectors corresponding to the values taken by the pulse direction $\OM'(t)$ are shown in figure \ref{fig:dOm-pt-amplitude-spinecho}(a).
The error integral vectors $\pp_i$ are directed along the three vectorial values taken by the error $\OM'_1(t)$. The resulting three-step walk is shown in figure \ref{fig:dOm-pt-amplitude-spinecho}(b). Note that the vector for step II has twice the length of that of steps I and II because it is a full $\pi$-pulse rather than a $\pi/2$-pulse. The three-step walk is not closed so the first-order constraint is not obeyed.

\begin{figure}[t]
	\includegraphics[scale=1]{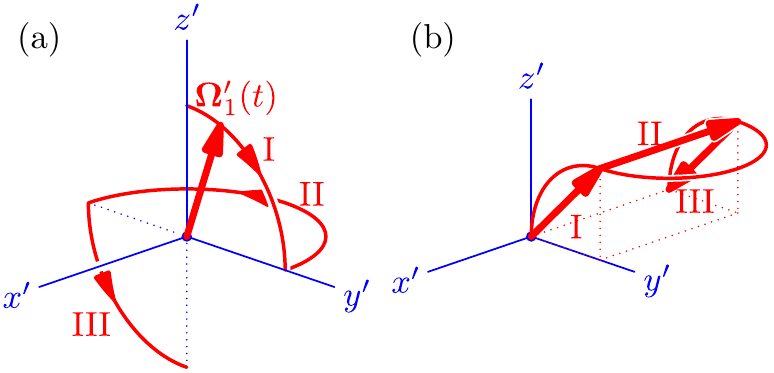}
    \caption{(a): the detuning error $\OM_1'(t)$ for the spin-echo sequence $R_0^{\pi/2}\rightarrow R_{\pi/2}^{\pi}\rightarrow R_0^{\pi/2}$ in the toggling frame. The three sections of the sequence are again labeled  I, II, and III. The arcs have radius $\Delta$. (b): error integral $\pp(t)$ and three error integral vectors $\pp_i$ for the spin-echo sequence with detuning error.}
    \label{fig:dOm-pt-detuning-spinecho}
\end{figure}

For the case of the detuning error, $\OM_1$ is constant in the lab frame, pointing in $z$ with magnitude $\Delta$, but in the toggling frame the direction of $\OM_1'(t)=\mathbf{\hat R}^{-1}\OM_1$ rotates at a constant rate. This apparent motion is the result of the rotation of the toggling frame and is shown in figure \ref{fig:dOm-pt-detuning-spinecho}.
Note that the figure \ref{fig:dOm-pt-detuning-spinecho}(a) resembles the rotation of the Bloch vector on the Bloch sphere, but should not be mistaken as such. Treating $\pp(t)$ as the motion of a phantom particle as mentioned previously, we can deduce that since its velocity $\OM_1'(t)$ rotates at a constant rate, the phantom particle's trajectory is composed of circular arcs, as shown in figure \ref{fig:dOm-pt-detuning-spinecho}. Step II is a semi-circle while steps I and III are quarter-circles. The three-step walk in the presence of detuning is also not closed, so the spin-echo sequence is not fully-compensating with respect to either amplitude errors or detuning errors.

For the case of amplitude error, the first-order error is suppressed provided the initial orientation of the Bloch vector is along the $z$-axis because the cross-product of Eq.(\ref{first-order-final}) is zero. This result can be understood as follows. Suppose that initially the Bloch vector points along the $z$-axis. The strength of the AC control pulse is too large, so the Bloch vector slightly overshoots the $-y$ direction during the first $\pi/2$ pulse around the $x$-axis. To first-order, a small error is generated in the $-z$ direction. The $\pi$-pulse along the $y$-axis flips this error around, causing the Bloch vector to slightly undershoot the $-y$ direction. The $\pi$-pulse acts here as a form of ``time-reversal", allowing the third pulse to undo the error caused by the first. This time-reversal is evident in figure \ref{fig:dOm-pt-amplitude-spinecho}, as $\pp_{\mathrm{I}}$ is anti-parallel with $\pp_{\mathrm{III}}$. A similar argument appealing to spatial reasoning may be made for the case of the detuning error. For detuning error, the first-order constraint is again obeyed if the initial spin orientation is along the $x$-axis.

\subsection{Three-step Sequences.}
As our second example we will construct three-step sequences that, unlike the spin-echo sequence, are fully-compensating. We restrict ourselves here to sequences of $\pi$-pulses because in that case the geometrical constructions ($N$-step walk) can be performed in the horizontal plane. We will ``reverse engineer" the sequence, i.e., we first construct closed error trajectories in the toggling frame and then work backwards to find the corresponding sequence in the toggling and lab frames. 

\begin{figure}[t]
    \includegraphics[scale=1]{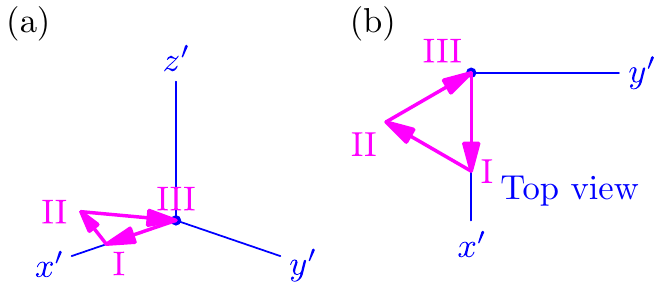}
    \caption{(a): the three error-integral vectors for a three-step sequence (equal pulse area for each pulse) that is fully-compensating to first-order for either kind of error must form an equilateral triangle. The labels I, II, and III refer to the three steps of the sequence. The triangle can be rotated and the direction of the arrows can be inverted. (b): top view.}
    \label{fig:pt-amplitude-threestep}
\end{figure}

With each of the three steps of the sequence, there is an associated error integral vector $\pp_i$ with $i=\mathrm{I,II,III}$. The error integral vectors are now restricted to the $xy$-plane because the pulses are rotations by $\pi$, and they have the same magnitude because the three pulses have the same duration. To produce the closed $N$-step walk of a fully-compensating sequence there is only one option: the three vectors must form an equilateral triangle (see figure \ref{fig:pt-amplitude-threestep}).
The next steps are finding the corresponding error vectors $\OM_1'(t)$ -- still in the toggling frame -- and then reconstructing the corresponding sequence in the lab frame. 

\begin{figure}[t]
	\includegraphics[scale=1]{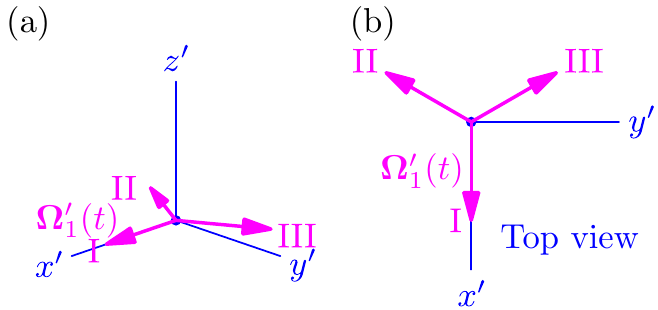}
    \caption{(a): the three vectorial values taken by the amplitude error $\OM_1'(t)$ for the three-step sequence. The labels I, II, and III refer to the three steps of the sequence. Since $\OM_1'(t)=\epsilon\OM'(t)$, we can immediately read off the pulse directions in the toggling frame, which form the following angles with $x'$-axis: $0\rightarrow-2\pi/3\rightarrow2\pi/3$.  (b): top view.}
    \label{fig:dOm-amplitude-threestep}
\end{figure}

The simplest case is again that of the amplitude error. Figure \ref{fig:dOm-amplitude-threestep} shows the three vectorial values that the error $\OM_1'(t)$ must take during the three steps, pointing from the center symmetrically to the vertices of an equilateral triangle. Since the error $\OM_1'(t)$ is parallel to $\OM'(t)$, we can immediately read off the pulse directions $\OM'(t)$ in the toggling frame. It must form the following angles with the $x'$-axis during the three steps respectively: $0\rightarrow-2\pi/3\rightarrow2\pi/3$. For the case of detuning error, special care needs to be taken in deducing the relationship between the pulse directions $\OM'(t)$ and $\pp_i$. This is discussed in detail in appendix \ref{sec.detuning-first-order}, the key conclusion of which is: for detuning error, the direction of $\pp_i$ for an odd-numbered step is a 90$^\circ$ counter-clockwise rotation from its pulse direction; for an even-numbered step, clockwise.

We are now ready for the final step: the reconstruction of the sequence in the lab frame. Denote this unknown sequence by 
\begin{equation}
    R_{\phi_1}^{\pi}\rightarrow R_{\phi_2}^{\pi}\rightarrow R_{\phi_3}^{\pi}.
\end{equation}
For each $\pi$-pulse, the lab frame $z$-axis ``toggles" between pointing along the $z'$ and $-z'$ directions in the toggling frame. Therefore, the pulse directions, which lie in the $xy$-plane in the lab frame, lie in the $x'y'$-plane in the toggling frame. A formula shown by \cite{wimperis_broadband_1994} relates rotation angles in the lab and toggling frames,
\begin{equation}
    \phi_j'=-(-1)^j\phi_j-\sum_{k=1}^{j-1}(-1)^k2\phi_k,
    \label{toggle-phases-eq}
\end{equation}
where $\phi_k={\phi_1,\phi_2,\phi_3,...}$ specify the pulse directions in the lab frame, and $\phi_j'$ are the corresponding directions in the toggling frame. This transformation is also its own inverse. One can work in the toggling frame and convert the angles back to the lab frame if needed. We first apply this transformation to the case of amplitude error. In the toggling frame, following our previous discussion, the angles between the pulse directions and the $x'$-axis are $0\rightarrow -2\pi/3\rightarrow 2\pi/3$ (Figure \ref{fig:dOm-amplitude-threestep}). Converting back to the lab frame angles, using Eq.\eqref{toggle-phases-eq}, produces the sequence: 
\begin{equation}
    R_0^{\pi}\rightarrow R_{2\pi/3}^{\pi}\rightarrow R_0^{\pi}.
    \label{threestep-spec-amplitude}
\end{equation}
A constant can of course be added to all three angles without affecting first-order error suppression. 

Next, for the case of detuning error, we have shown previously that the pulse direction $\OM'(t)$ needs to form the following three angles with the $x'$-axis: $0\rightarrow -\pi/3\rightarrow -2\pi/3$ (Figure \ref{fig:dOm-detuning-threestep}). Using the transformation formula Eq.\eqref{toggle-phases-eq} produces the composite pulse: 
\begin{equation}
    R_0^{\pi}\rightarrow R_{\pi/3}^{\pi}\rightarrow R_0^{\pi}.
    \label{threestep-spec-detuning}
\end{equation}

\subsection{Comparison between the Quasi-Classical and Time-Evolution Operator Frameworks (I).}
In ref.\cite{jones_designing_2013, merrill_transformed_2014} the time-evolution operator method is used to derive constraints for a sequence of $\pi$-pulses to provide universal error correction to first-order. The constraints on the orientation angles of the pulse fields are
\begin{align}
\sum_j\sigma_{\phi_j'}&=0\ \ \mathrm{(amplitude)}, \label{jones-1-a}\\
\sum_j\sigma_{\phi_j''}&=0\ \ \mathrm{(detuning)}, \label{jones-1-b}
\end{align}
where
\begin{align}
\phi_j'&=(-1)^{j+1}\phi_j+\sum_{k<j}(-1)^{k+1}2\phi_k,
\label{jones-2-a}\\
\phi_j''&=\phi_j'+(-1)^{j+1}\pi/2,
\label{jones-2-b}
\end{align}
and where the $\sigma_{\phi}=\cos\phi\ \sigma_x+\sin\phi\ \sigma_y$ are linear combinations of Pauli matrices. Note that this last relation already suggests a link with vectors in the $xy$-plane. 

To establish a connection between the two frameworks, we first rewrite the approach described in the previous section in a more mathematical language where we no longer restrict ourselves to three-step sequences (the sequence can be composed of any odd number of $\pi$-pulses). The key relationships that we illustrated with the three-step sequences can be shown to remain the same. The following algorithm makes precise the relationship between the pulse direction $\OM(t)$ in the lab frame, the pulse direction $\OM'(t)$ in the toggling frame, the $N$-step walk for amplitude error, the $N$-step walk for detuning error, and the conditions for error-suppression in case of either type of error.
\begin{enumerate}
    \item Use Eq.\eqref{toggle-phases-eq} to transform from the lab frame pulse directions ($\phi_k$) to the toggling frame pulse directions ($\phi'_k$).
    \item Draw the toggling frame pulse directions as unit length vectors laying in the $x'y'$-plane. The complex number notation can make this statement more precise: draw the following set of complex numbers as vectors on the complex plane, $\{\mathrm{e}^{\mathrm{i}\phi'_1},\mathrm{e}^{\mathrm{i}\phi'_2},\mathrm{e}^{\mathrm{i}\phi'_3}...\}$. Label the step number for each vector.
    \item \begin{enumerate}
        \item To obtain the $N$-step walk made up of amplitude error integral vectors $\pp_i$: starting from the origin, parallel transport the vectors so that they connect tip-to-tail in order. The final result is the $N$-step walk made from amplitude error integral vectors $\pp_i$. In terms of complex numbers this can be expressed as:
        \begin{equation}
            \mathrm{e}^{\mathrm{i}\phi'_1}+\mathrm{e}^{\mathrm{i}\phi'_2}+\mathrm{e}^{\mathrm{i}\phi'_3}+...
            \label{amplitude-pt-with-complex-numbers}
        \end{equation}
        \item To obtain the $N$-step walk made up of detuning error integral vectors $\pp_i$: return to step 2. Rotate those vectors for odd steps counterclockwise by $\pi/2$ and rotate those for the even steps clockwise by $\pi/2$. Starting from the origin, parallel transport the resulting vectors so that they connect tip-to-tail in order. The final result is the $N$-step walk made from detuning error integral vectors $\pp_i$. In terms of complex numbers this can be expressed as:
        \begin{equation}
            \mathrm{i}\mathrm{e}^{\mathrm{i}\phi'_1}-\mathrm{i}\mathrm{e}^{\mathrm{i}\phi'_2}+\mathrm{i}\mathrm{e}^{\mathrm{i}\phi'_3}-...
            \label{detuning-pt-with-complex-numbers}
        \end{equation}
        Mark the odd steps with ``$+$" signs and even steps with ``$-$" signs on paper.
    \end{enumerate}
    \item The sequence is fully-compensating to first-order with respect to amplitude error if the figure formed in Eq.\eqref{amplitude-pt-with-complex-numbers} is a closed polygon, and it is fully-compensating to first-order with respect to detuning error if the figure formed in Eq.\eqref{detuning-pt-with-complex-numbers} is a closed polygon. In terms of the complex numbers
    \begin{align}
    	&\sum_{j}\mathrm{e}^{\mathrm{i}\phi_j'} \label{amplitude-pt-with-complex-numbers-2}=0   \ \ \ \ (\mathrm{amplitude}),\\&
    	\sum_{j}(-1)^{j-1}\mathrm{i}\mathrm{e}^{\mathrm{i}\phi_j'} \label{detuning-pt-with-complex-numbers-2}=0 \ \ \ \ (\mathrm{detuning}).
    \end{align}
\end{enumerate}
It is easy to verify this algorithm for the three-step sequences of the previous section. Because the procedure is invertible one can reverse-engineer a desirable $N$-step walk to get the specification of the composite pulse, as explicitly shown for three-step sequences. 

Next, expand the constraints expressed in complex exponential form in Eq.\eqref{amplitude-pt-with-complex-numbers-2} and \eqref{detuning-pt-with-complex-numbers-2} into real and imaginary parts, perform the substitution $1\rightarrow\sigma_x$, $\mathrm{i}\rightarrow\sigma_y$, and then use the definition of $\phi_j''$ from Eq.\eqref{jones-2-b} where appropriate. One obtains
    \begin{align*}
    	0&=\sum_{j}\mathrm{e}^{\mathrm{i}\phi_j'}=\sum_{j}\cos\phi_j'+\mathrm{i}\sin\phi_j'\\
    	&\Rightarrow\mathrm{(substitution)}\\
    	0&=\sum_{j}\cos\phi_j'\ \sigma_x+\sin\phi_j'\ \sigma_y\\
    	&=\sum_{j}\sigma_{\phi_j'}\ \mathrm{(amplitude\ error)},
    \end{align*}
    and,
    \begin{align*}
    	0&=\sum_{j}(-1)^{j-1}\mathrm{i}\mathrm{e}^{\mathrm{i}\phi_j'}=\sum_{j}\mathrm{e}^{\mathrm{i}(\phi_j'+(-1)^{j-1}\pi/2)}\\
    	&=\sum_{j}\mathrm{e}^{\mathrm{i}\phi_j''}=\sum_{j}\cos\phi_j''+\mathrm{i}\sin\phi_j''\\
    	&\Rightarrow\mathrm{(substitution)}\\
    	0&=\sum_{j}\cos\phi_j''\ \sigma_x+\sin\phi_j''\ \sigma_y\\
    	0&=\sum_{j}\sigma_{\phi_j''}\ \mathrm{(detuning\ error)}.
    \end{align*}
   This reproduces Eqs.\eqref{jones-1-a} and \eqref{jones-1-b}. It follows that, at least for the first-order constraint, the two frameworks produce precisely the same result. Appendix \ref{sec.five-step-sequences} demonstrates how to apply the constraints formulated in this section to five-step sequences (such as the Knill sequence) that are fully-compensating to first order in both amplitude and detuning error.

\section{Second-Order Constraint.}
Second-order errors are eliminated when Eq.\eqref{second-order} is satisfied. Assume that the first-order constraint is obeyed and substitute the definition of the error integral $\pp(t)$ (see Eq.\eqref{pt-definition}) into the second-order constraint,
\begin{align}
    \rr_2'|_{t=t_{\mathrm{f}}}&=\int_0^{t_{\mathrm{f}}}\diff{}{s}\pp(s)\times(\pp(s)\times\rr_0(0))\dd s = 0.
\end{align}
Using the fact that the $\pp(t)$ curve is closed, this equation can be manipulated (see appendix \ref{sec.second-order}) into the form:
\begin{equation}
    \rr_2'|_{t=t_{\mathrm{f}}}=\rr_0(0)\times\left(\frac{1}{2}\oint\pp\times\dd\pp\right) = 0.
    \label{second-order-manip}
\end{equation}
The \textit{vector area} $\bf{A}=\int_{\mathcal{S}}\dd\mathbf{S}$ of any surface bounded by the same curve $\mathcal{C}$ is given by
\begin{equation}
   {\bf{A}}= \frac{1}{2}\oint_{\mathcal{C}}{\bf{r}}\times d\bf{r},
\end{equation}
where ${\bf{r}}$ traces out the boundary curve \cite{griffiths_introduction_2014}. It follows that if the vector area of any surface enclosed by the error integral is zero, then the second-order error is zero independent of the initial condition. The corresponding pulse is then fully-compensating to second-order. Otherwise, the sequence suppresses second-order error only for states initially polarized in the same direction as the vector area. 

\subsection{Three-step Sequences.}
First apply the second-order constraint to the two three-step sequences Eq.\eqref{threestep-spec-amplitude} and Eq.\eqref{threestep-spec-detuning}, which were shown to be fully-compensating to first-order for, respectively, amplitude and detuning error. Start with Eq.\eqref{threestep-spec-amplitude}. The signed area enclosed by the vectors I, II, III is an equilateral triangle with a normal vector along the $z'$ direction. Thus this sequence is \textit{not} fully-compensating to second-order. Note though that the second-order error does vanish for initial states polarized along the $z$-axis. 

\begin{figure}[t]
	\includegraphics[scale=1]{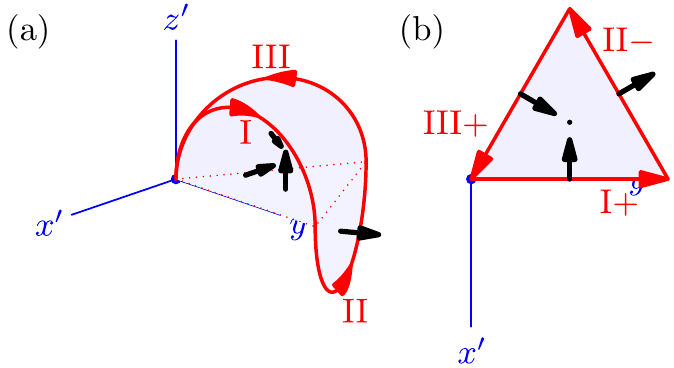}
    \caption{(a): decomposition of the vector area $\frac{1}{2}\oint\pp\times\dd\pp$ enclosed by the detuning error integral $\pp(t)$ for the three-step sequence of Eq.\eqref{threestep-spec-detuning}. The vector areas of each of the four sub-areas are indicated by small black arrows. (b): the three vector areas associated with semi-circle trajectories. A plus sign indicates the semi-circle is above the $z=0$ plane and a minus sign below the $z=0$ plane. The vector sum from the semi-circles lies in the $x'y'$-plane and cannot cancel the vector area of the enclosed triangle, which lies along the $z'$-axis.}
    \label{fig:pt-detuning-threestep-second-order}
\end{figure}

Next, consider the sequence Eq.\eqref{threestep-spec-detuning} for the case of detuning error. The enclosed area can be divided into three planar parts bounded by semi-circles plus a planar part bounded by the triangle, as shown in figure \ref{fig:pt-detuning-threestep-second-order}. The directions of vector areas of the three semi-circles are all in the $x'y'$-plane and thus cannot cancel the vector area of the triangle, which is along the $z'$ direction.  It follows that the net vector area is non-zero as well, so this sequence is also \textit{not} fully-compensating to second-order with respect to detuning error. Examples of sequences that are fully-compensating to second order are provided in appendix \ref{sec.five-step-sequences}.
 
\subsection{Comparison between the Quasi-Classical and Time-Evolution Operator Frameworks (II).}
Following \cite{jones_designing_2013} the condition for second-order amplitude error suppression in the time-evolution operator method is
\begin{equation}
\sum_j\sum_{k<j}[\sigma_{\phi_j'},\sigma_{\phi_k'}]=0\ \mathrm{(amplitude)}.
\end{equation}
As discussed in Ref. \cite{merrill_transformed_2014} (see Eq.(3)) the sum of commutators in this equation is in fact the signed area formed by a walk defined by the different angles in the plane. This can be seen by rewriting the condition as
\begin{equation}
\sum_j\sum_{k<j}\sin(\phi_j'-\phi_k')\sigma_z=0,
\end{equation}
and recalling that the cross-product of two unit vectors is the vector area of the parallelogram spanned by them. Similarly, the condition for the second-order detuning error is
\begin{equation}
\sum_j\sum_{k<j}\sin(\phi_j''-\phi_k'')\sigma_z=0 \ \mathrm{(detuning)}.
\label{jones-3-b}
\end{equation}
For both the amplitude and detuning errors, the signed area formed by the walk in the plane specified by the different angles must be zero. We conclude that up to second order the quasi-classical framework produces the same error-suppressing pulse sequences as the time-evolution operator method.

\section{General Rotation Angles.}
\label{sec.theta-pulse}
We have assumed so far that the desired net effect of the sequence is a $\pi$-pulse. The quasi-classical framework for amplitude error suppression can be extended if the net effect of the sequence is a rotation around the $x$-axis by some arbitrary angle $\theta$. We may choose to set up the pulse sequence as follows: $R_{\phi_1}^{\pi}\to R_{\phi_2}^{\pi}\to R_{\phi_3}^{\pi}\to R_{\phi_4}^{\pi}\to R_0^{\theta}$. For the overall net effect to be correct, we let the first 4 pulses have the net effect of identity. It is most straightforward, though not necessary, to use the propagator framework to match the net effect:
\begin{align}
	\hat{U}_0&=\exp[-i\pi\hat\sigma_{\phi_4}/2]\cdots\exp[-i\pi\hat\sigma_{\phi_1}/2] \\
	&=\hat\sigma_{\phi_4}\hat\sigma_{\phi_3}\hat\sigma_{\phi_2}\hat\sigma_{\phi_1} \\
	&=\exp[-i(\phi_4-\phi_3+\phi_2-\phi_1)\hat\sigma_z]\\
	&=\cos(\phi_4-\phi_3+\phi_2-\phi_1)\mathbf{I} \nonumber \\
	&\quad\quad-i\sin(\phi_4-\phi_3+\phi_2-\phi_1)\hat\sigma_z
\end{align}
Thus we need
\begin{align}
	\phi_4-\phi_3+\phi_2-\phi_1 = 2\pi k,
\end{align}
where $k$ is an integer. Since we are free to add/subtract $2\pi$ to/from the phases, we may simply write the constraint as
\begin{equation}
	\phi_1-\phi_2+\phi_3-\phi_4=0
\end{equation}
Now work in the toggling frame and rewrite the above net effect constraint with the toggling frame phases,
\begin{multline}
	\phi_1'-(2\phi_1'-\phi_2')+(2\phi_1'-2\phi_2'+\phi_3)\\-(2\phi_1'-2\phi_2'+2\phi_3'-\phi_4')=0.
\end{multline}
Therefore,
\begin{align}
	-\phi_1'+\phi_2'-\phi_3'+\phi_4'&=0.
\end{align}
Under this condition, the fifth toggling frame phase is
\begin{align}
	\phi_5' &= 2\phi_1-2\phi_2+2\phi_3-2\phi_4+\phi_5 \\
	&=2(\phi_1-\phi_2+\phi_3-\phi_4)+0 \\
	&=0.
\end{align}
Now define $\alpha, \beta, \gamma, \delta$ such that
\begin{align}
	\phi_1' &= \alpha + \gamma, \phi_2' = \alpha - \gamma, \\
	\phi_3' &= \beta + \delta, \phi_4' = \beta - \delta.
\end{align}
Plug these definitions into the net effect constraint gets us
\begin{align}
	0 &= (\alpha + \gamma) - (\alpha - \gamma) + (\beta + \delta) - (\beta - \delta) \\
	&= 2\gamma + 2\delta 
\end{align}
Hence we must have $\delta = -\gamma$. So far the results arise from the net effect constraint only. 

Since the final pulse is not a $\pi$ pulse, the usual equilateral pentagon that determines first-order error suppression will have unequal side-lengths, namely $1, 1, 1, 1, \theta/\pi$ (disregarding overall scaling) where the last side-length is shortened (or lengthened) due to the specified pulse area. Next we force the polygon to be closed in the toggling frame.
\begin{figure}[t]
    \includegraphics[scale=0.25]{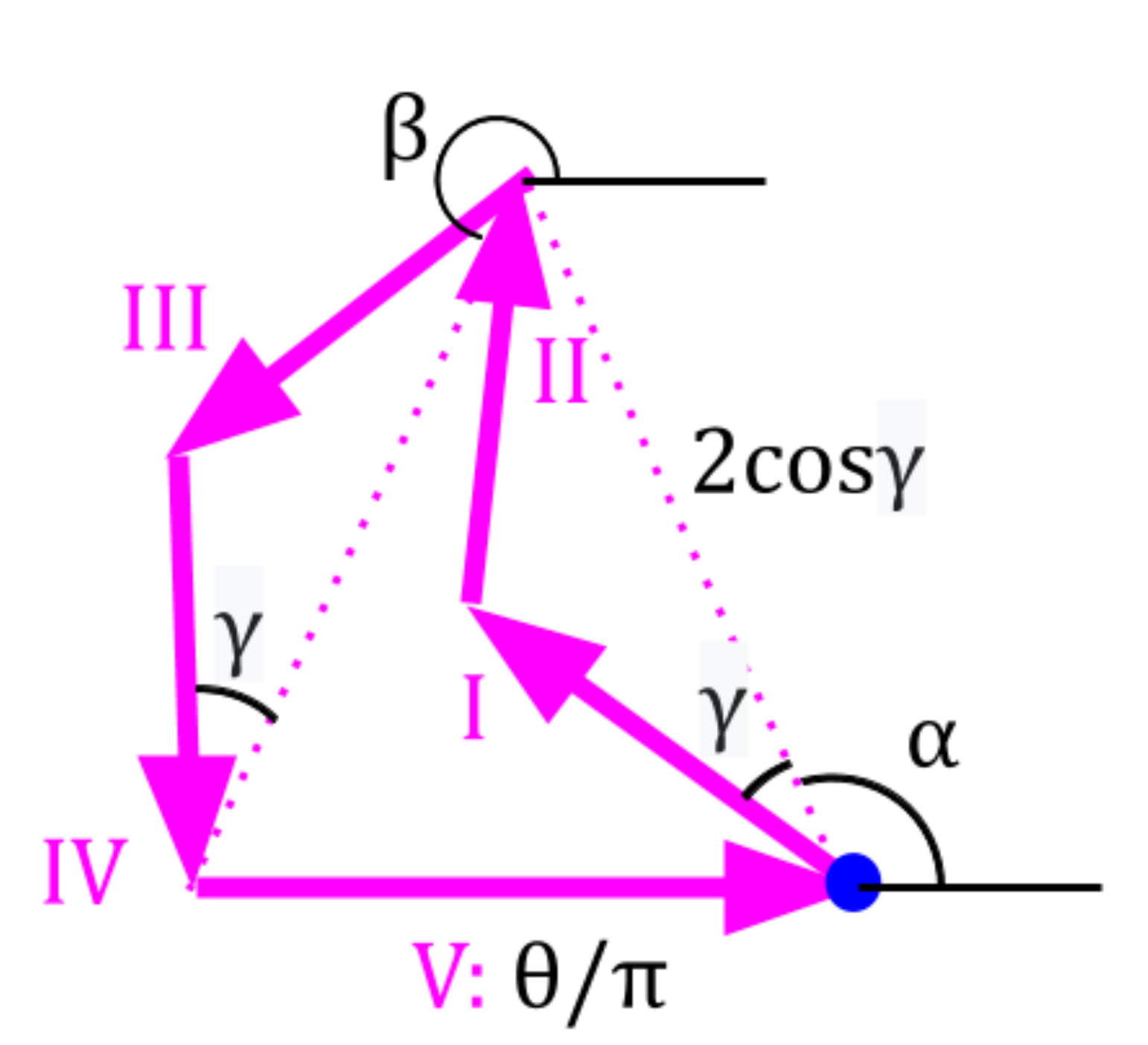}
    \caption{Amplitude error integral vectors for a five step sequence with net effect of a $\theta$ rotation around the $x$ axis.}
    \label{fig:five-step-two-param}
\end{figure}
Figure \ref{fig:five-step-two-param} shows what shape the error integral vectors must be given the net effect constraint and first-order amplitude error suppression. The auxiliary lines reveal three isosceles triangles. Because the error integral vectors for step I, II, III, and IV have length 1 and the base angles are $\gamma$, both auxiliary lines have length $2\cos\gamma$. The relationship between the length of the auxiliary lines and the length of step V can also be determined from trigonometry, which produces
\begin{equation}
	\cos\gamma\cos\alpha=-\frac{\theta}{4\pi}
\end{equation}
Now we can express the lab frame phases in terms of $\alpha$ and $\gamma$
\begin{equation}
	\boldsymbol{\phi}=\alpha(1,1,-1,-1,0)+\gamma(1,3,3,1)
\end{equation}
It is a one-parameter family of sequences which accomplish a $\theta$ rotation around the $x$-axis and suppresses first-order amplitude error. It is obvious from Figure \ref{fig:five-step-two-param} that when $\alpha=\pi$ (which in fact reproduces the $\mathrm{BB}_1(\theta)$ sequence \cite{wimperis_broadband_1994}), we get a polygon with zero vector area, so the sequence suppresses error to second order.

\section{Formal Connection Between the Quasi-Classical and Propagator Frameworks.}
\label{sec.gen-rot-ang}
Let $\hat{U}(t)$ be the time-evolution operator of the spin wavefunction (or propagator from hereon). It obeys the equation 
\begin{align}
	\mathrm{i}\frac{\mathrm{d}\hat{U}(t)}{\mathrm{d}t} = \left[(\OM(t)+\OM_1(t))\cdot{\hat{\boldsymbol{\sigma}}}/2\right]\hat{U}(t). \label{prop-eq}
\end{align}
Transform to the toggling frame by decomposing $\hat{U}(t)$ as
\begin{align}
	\hat{U}(t) = \hat{U}_0(t)\hat{V}(t), \label{prop-decompose}
\end{align}
where $\hat{U}_0$ will be called the error-free propagator, and $\hat{V}$ the error propagator. Here, $\hat{U}_0(t)$ satisfies
\begin{align}
	\mathrm{i}\frac{\mathrm{d}\hat{U}_0(t)}{\mathrm{d}t} = \left(\OM(t)\cdot{\hat{\boldsymbol{\sigma}}}/2\right)\hat{U}(t).
\end{align}
Assuming a sequence that accomplishes the desired net effect in the error-free case, we are only concerned about $V(t)$. Substituting Eq.\eqref{prop-decompose} into Eq.\eqref{prop-eq} gives us an equation for $V(t)$,
\begin{align}
	\frac{\mathrm{d}\hat{V}(t)}{\mathrm{d}t} = -\mathrm{i}\left(\OM_1'(t)\cdot{\hat{\boldsymbol{\sigma}}}/2\right)\hat{V}(t). \label{prop-eq-toggling}
\end{align}
We used the fact that $U_0^{\dagger}(t)\left(\OM_1(t)\cdot{\hat{\boldsymbol{\sigma}}}\right)U_0=\OM_1'(t)\cdot{\hat{\boldsymbol{\sigma}}}$, where $\OM_1'(t)$ is the amplitude or detuning error in the toggling frame. Since $\hat{V}$ is a unitary operator, it must have the form
\begin{align*}
	\hat{V} = \exp[-\mathrm{i}\mathbf{a}\cdot\boldsymbol\hat{\boldsymbol\sigma}/2],
\end{align*}
where $\mathbf{a}$ is a dimensionless 3D vector whose magnitude is a measure of the error. For operator or matrix equations of the form
\begin{align}
	\frac{\mathrm{d}\hat{V}(t)}{\mathrm{d}t} = \hat{A}(t)\hat{V}(t),
\end{align}
the Magnus expansion provides the solution,
\begin{align*}
	\hat{V}(t) &= \exp\left[\hat{\Phi}(t)\right], \\
	\text{where } & \hat{\Phi}(t) = \hat{\Phi}_1(t) + \hat{\Phi}_2(t) + \cdots\text{ and} \\
	\hat{\Phi}_1(t) &= \int_0^t \hat{A}(t_1)\mathrm{d}t_1, \\
	\hat{\Phi}_2(t) &= \frac{1}{2}\int_0^t\mathrm{d}t_1\int_0^{t_1}\mathrm{d}t_2\ [\hat{A}(t_1), \hat{A}(t_2)], \\
	&\cdots
\end{align*}
Higher-order terms in this series are integrals of increasingly longer nested commutators of $\hat{A}(t)$. Setting $\hat{A}(t)=-\mathrm{i}\OM_1'(t)\cdot\hat{\boldsymbol\sigma}/2$ gives
\begin{align}
	\hat{\Phi}_1(t) &= -\mathrm{i}\left(\int_0^t \OM_1'(t_1)\mathrm{d}t_1\right)\cdot\hat{\boldsymbol\sigma}/2 \\
	&= -\mathrm{i}\ \pp(t)\cdot\hat{\sigma}/2 \label{magnus_1} \\
	\hat{\Phi}_2(t) &= -\frac{1}{8}\int_0^t\mathrm{d}t_1\int_0^{t_1}\mathrm{d}t_2\ [\OM_1'(t_1)\cdot\hat{\sigma}, \OM_1'(t_2)\cdot\hat{\boldsymbol\sigma}] \\
	&= -\mathrm{i}\left(\frac{1}{2}\int_0^t\mathrm{d}t_1\int_0^{t_1}\mathrm{d}t_2\ \OM_1'(t_1)\times\OM_1'(t_2)\right)\cdot\hat{\boldsymbol\sigma}/2 \\
	&= -\mathrm{i}\left(\frac{1}{2}\int_0^t\mathrm{d}t_1\ \pp(t_1)\times\frac{\mathrm{d}\pp(t_1)}{\mathrm{d}t}\right)\cdot\hat{\boldsymbol\sigma}/2 \\
	&= -\mathrm{i}\left(\frac{1}{2}\int_0^t\pp\times\mathrm{d}\pp\right)\cdot\hat{\boldsymbol\sigma}/2, \label{magnus_2}
\end{align}
where we made use of the identity
\begin{equation}
	[-\mathrm{i}\mathbf{A}\cdot\hat{\boldsymbol\sigma}/2,-\mathrm{i}\mathbf{B}\cdot\hat{\boldsymbol\sigma}/2]=-\mathrm{i}(\mathbf{A}\times\mathbf{B})\cdot\hat{\boldsymbol\sigma}/2. \label{commute-id}
\end{equation}
This identity is the key relation between the commutators that appear in the propagator framework and the cross-products that appear in the quasi-classical framework. Equations \eqref{magnus_1} and \eqref{magnus_2} are the first- and second-order constraints that we derived previously within the classical framework. The Magnus expansion of the propagator leads to the same first- and second-order error suppression as the quasi-classical framework. We conjecture that the correspondence between the quasi-classical and propagator frameworks holds to all orders.

\section{Conclusion.}
In conclusion, we demonstrated that composite pulses that suppress systematic amplitude and detuning errors can be derived from the quasi-classical equation for the spin expectation value. This suggests that when teaching a class in quantum computing, error correction by composite pulses can be discussed most easily in terms of classical physics with simple diagrams. It would be interesting to investigate whether this conclusion will continue to hold for two entangled qubits. An objection could be that loss of phase coherence due to interaction with a dissipative environment cannot be treated in a completely classical context. It should be kept in mind that the classical Bloch equations do account for loss of phase coherence and we conjecture that the design of error-suppressing pulse sequences in the presence of dissipative coupling to the environment can be treated satisfactorily within the framework of the Bloch equations. It is interesting to speculate about error correction methods for quantum computing that are not quasi-classical in nature. Topological quantum computing \cite{raussendorf_topological_2007} is an obvious possibility but it is not clear how a topologically protected state would behave in the presence of systematic detuning and amplitude errors.

The focus of this paper has been mostly on sequences with an odd number of $\pi$-pulses. The technique becomes less straightforward for the case of constructing error-suppressing sequences from $\pi/2$-pulses because closed walks of $\pp_i$ vectors would no longer be planar. This is not a prohibitive objection and studying sequences of $\pi/2$ pulses is nevertheless a promising direction because the extra spatial degree of freedom could allow for shorter sequences that give the same order of error suppression. A related, but distinct, direction for potential development is to apply it to sequences that accomplish a net $\pi/2$-pulse (or any other fractional-$\pi$ pulses) rather than just a $\pi$-pulse. One could prepend or append any number of correcting $\pi$- or 2$\pi$-pulses to the target pulse. The bulk of the spin evolution would still produce $\pp(t)$ curves that are planar so the technique developed in this paper would still be applicable. A partial solution to this problem is outlined in appendix \ref{sec.theta-pulse}.

Another important assumption was that the amplitude or detuning errors are constant in time. In actuality, a drift over time often occurs. To model the drift in the experimental parameters one could allow for different errors for each step in the sequence, i.e. a set of $\{\epsilon_1,\epsilon_2,\epsilon_3...\}$ or $\{\delta_1,\delta_2,\delta_3...\}$. The steps of the error integral vectors would no longer be of equal length in that case. The solution of the constraints would follow the same geometric principles, but it would be mathematically more involved.

Acknowledgements:  I would like to thank R. Bruinsma, W. Campbell and C. Romes for guidance, insights, and discussions and R. Bruinsma for assistance in the preparation of the manuscript. Initial work on this project was carried out by Xingchen Fan and Clementine Domine with support from the NSF-DMR under CMMT Grant 1836404.

\appendix
\section{Rotating Wave Approximation and the Classical Equation of Motion.}
\label{sec.setup-math}
Begin with the \textit{lab-frame} Hamiltonian
\begin{equation}
    H=\hbar/2[\omega_0\hat\sigma_z+\Omega\cos(\omega t+\phi)\hat\sigma_x+\Omega\sin(\omega t+\phi)\hat\sigma_y],
\end{equation}
which describes the rotation of the spin under the DC and AC magnetic field. Following \cite{vandersypen_nmr_2005}, cancel the spin precession caused by the DC field by transforming into a rotating frame. Define
\begin{equation}
    \ket{\psi}=U(t)\ket{\psi}^{\mathrm{rot}},
\end{equation}
where $U=\exp\left(\frac{-i\omega t}{2}\hat\sigma_z\right)$ is a rotation about the $z$-axis. Plugging in this definition into the Schrodinger's equation $\mathrm{i}\partial_t\ket{\psi}=H\ket{\psi}$ produces the effective Hamiltonian
\begin{equation}
    H'=U^\dagger HU-\mathrm{i}\hbar U^\dagger \diff{U}{t}.
\end{equation}
Notice that
\begin{equation}
    \cos\theta\ \hat\sigma_x+\sin\theta\ \hat\sigma_y = \mathrm{e}^{-\mathrm{i}\theta\hat\sigma_z}\hat\sigma_x = \hat\sigma_x\mathrm{e}^{\mathrm{i}\theta\hat\sigma_z}. \label{Pauli-trick}
\end{equation}
Hence
\begin{align}
    &\ U^\dagger (\cos(\omega t+\phi)\ \hat\sigma_x+\sin(\omega t+\phi)\ \hat\sigma_y) U    \nonumber\\
    &=\mathrm{e}^{\frac{\mathrm{i}\omega t}{2}\hat\sigma_z} \left(\mathrm{e}^{-\mathrm{i}(\omega t+\phi)\hat\sigma_z}\hat\sigma_x\right)\mathrm{e}^{\frac{-\mathrm{i}\omega t}{2}\hat\sigma_z}   \nonumber\\
    &=\mathrm{e}^{\frac{\mathrm{i}\omega t}{2}\hat\sigma_z}\mathrm{e}^{-\mathrm{i}(\omega t+\phi)\hat\sigma_z}\mathrm{e}^{\frac{\mathrm{i}\omega t}{2}\hat\sigma_z}\hat\sigma_x    \nonumber\\
    &=\mathrm{e}^{-\mathrm{i}\phi\hat\sigma_z}\hat\sigma_x    \nonumber\\
    &=\cos\phi\ \hat\sigma_x+\sin\phi\ \hat\sigma_y.
\end{align}
Because $\hat{U}$ commutes with $\hat\sigma_z$ and $U^\dagger\diff{U}{t}=-\mathrm{i}\omega\hat\sigma_z/2$, the effective Hamiltonian is
\begin{equation}
    H'=\hbar/2[(\omega_0-\omega)\hat\sigma_z+\Omega\cos\phi\ \hat\sigma_x+\Omega\sin\phi\ \hat\sigma_y].
\end{equation}

In the case where the AC magnetic field is linearly polarized as opposed to circularly polarized, the field can be decomposed into two counter-rotating circularly polarized waves. The \text{lab-frame} Hamiltonian will have the term
\begin{align}
    H_{\mathrm{rf}}&=2\Omega\cos(\omega t+\phi)\hat\sigma_x\\
    &=\Omega(\cos(\omega t+\phi)\hat\sigma_x+\sin(\omega t+\phi)\hat\sigma_y)    \nonumber\\
    &\ \ \ +\Omega(\cos(\omega t+\phi)\hat\sigma_x-\sin(\omega t+\phi)\hat\sigma_y).
\end{align}
Under the unitary transformation $U^\dagger H_{\mathrm{rf}}U$, the second term above produces
\begin{align}
    &\ U^\dagger (\cos(\omega t+\phi)\ \hat\sigma_x-\sin(\omega t+\phi)\ \hat\sigma_y) U    \nonumber\\
    &=\mathrm{e}^{\frac{\mathrm{i}\omega t}{2}\hat\sigma_z} \left(\mathrm{e}^{\mathrm{i}(\omega t+\phi)\hat\sigma_z}\hat\sigma_x\right)\mathrm{e}^{\frac{-\mathrm{i}\omega t}{2}\hat\sigma_z}   \nonumber\\
    &=\mathrm{e}^{\frac{\mathrm{i}\omega t}{2}\hat\sigma_z}\mathrm{e}^{\mathrm{i}(\omega t+\phi)\hat\sigma_z}\mathrm{e}^{\frac{\mathrm{i}\omega t}{2}\hat\sigma_z}\hat\sigma_x    \nonumber\\
    &=\mathrm{e}^{\mathrm{i}(2\omega t+\phi)\hat\sigma_z}\hat\sigma_x   . \nonumber
\end{align}
If the drive frequency $\omega$ is close to the Larmor frequency $\omega_0$, then this high-frequency time-dependent term has only a minor effect and can be dropped. This dropping of the high frequency term is known as the rotating wave approximation. Hence the wave rotating against the Larmor precession caused by the DC field can be neglected.

Now define the vector $\OM=(\Delta,\Omega\cos\phi,\Omega\sin\phi)$, where, $\Delta=\omega_0-\omega$. Then the effective Hamiltonian in the rotating frame is
\begin{equation}
    H'=\frac{\hbar}{2}\OM\cdot\boldsymbol\sigma.
\end{equation}
By Ehrenfest Theorem, the evolution of the Bloch vector is
\begin{equation}
    \diff{\braket{\hat\sigma_k}}{t}=\left<\frac{i}{\hbar}[H',\hat\sigma_k]\right>+\left<\pdiff{\hat\sigma_k}{t}\right>.
\end{equation}
The time derivative of the operator vanishes. The commutator can be evaluated
\begin{equation}
    [H',\hat\sigma_k]=\frac{\hbar}{2}\Omega_j[\hat\sigma_j,\hat\sigma_k]=\frac{\hbar}{2}\Omega_j(2i\epsilon_{jkl}\hat\sigma_l)=-i\hbar\epsilon_{jlk}\Omega_j\hat\sigma_l.
\end{equation}
Therefore,
\begin{align}
    \diff{\braket{\hat\sigma_k}}{t}&=\epsilon_{jlk}\Omega_j\braket{\hat\sigma_l}  \nonumber\\
    \diff{\braket{\boldsymbol\sigma}}{t}&=\OM\times\braket{\boldsymbol\sigma}.
\end{align}

\section{Transformation into the Toggling Frame.}
\label{sec.toggling-frame}
This section is concerned with solving the equation
\begin{equation}
    \diff{\rr(t)}{t}=(\OM(t)+\OM_1(t))\times\rr(t)   , \label{master-eq-restated}
\end{equation}
by introducing the toggling frame controlled by the rotation operator $\mathbf{\hat R}$, which is the solution to the operator equation
\begin{equation}
    \diff{}{t}\mathbf{\hat R}(t)=\OM(t)\times\mathbf{\hat R}(t) ,\label{define-R}  
\end{equation}
with initial condition $\mathbf{\hat R}(0)=\mathbf{I}$. Let the motion of the Bloch vector be the composition of both the rotation of the toggling frame and the motion with respect to the toggling frame, i.e. $\rr(t)=\mathbf{\hat R}(t)\rr'(t)$, where $\rr'(t)$ is the Bloch vector in the toggling frame. Direct substitution into Eq. \eqref{master-eq-restated} gives
\begin{align}
    \diff{}{t}\left(\mathbf{\hat R}\rr'\right)&=(\OM+\OM_1)\times\left(\mathbf{\hat R}\rr'\right)\\
    \left(\diff{}{t}\mathbf{\hat R}\right)\rr'+\mathbf{\hat R}\diff{\rr'}{t}&=(\OM+\OM_1)\times\left(\mathbf{\hat R}\rr'\right),
\end{align}
where the time-dependence is suppressed in the notation. Using Eq.\eqref{define-R}, for the time derivative of $\mathbf{\hat R}$,
\begin{align}
    \OM\times\left(\mathbf{\hat R}\rr'\right)+\mathbf{\hat R}\diff{\rr'}{t}&=(\OM+\OM_1)\times\left(\mathbf{\hat R}\rr'\right).
\end{align}
The two terms involving $\OM$ can be cancelled,
\begin{align}
    \mathbf{\hat R}\diff{\rr'}{t}&=\OM_1\times\left(\mathbf{\hat R}\rr'\right).
\end{align}
Multiply by the inverse of the operator $\mathbf{\hat R}$ on both sides,
\begin{align}
    \diff{\rr'}{t}&=\mathbf{\hat R}^{-1}\left(\OM_1\times\left(\mathbf{\hat R}\rr'\right)\right).
\end{align}
If $\OM_1(t)=0$, then the time derivative of $\rr'(t)$ must vanish, and the Bloch vector merely rotates with the toggling frame, as should be expected in the absence of errors. If errors are present and $\OM_1(t)\neq0$, one can use the following identity to distribute the inverse rotation operator $\mathbf{\hat R}^{-1}$ into both operands of the cross product: 
\begin{align}
    \left(\mathbf{\hat R}\mathbf{A}\right)\times\left(\mathbf{\hat R}\mathbf{B}\right)=\mathbf{\hat R}\left(\mathbf{A}\times\mathbf{B}\right),
\end{align}
where $\mathbf{\hat R}$ is a proper rotation and $\mathbf{A}$, $\mathbf{B}$ are any pair of 3D vectors. Therefore
\begin{align}
    \diff{\rr'}{t}&=\mathbf{\hat R}^{-1}\left(\OM_1\times\left(\mathbf{\hat R}\rr'\right)\right)\\
    &=\left(\mathbf{\hat R}^{-1}\OM_1\right)\times\left(\mathbf{\hat R}^{-1}\mathbf{\hat R}\rr'\right)\\
    &=\left(\mathbf{\hat R}^{-1}\OM_1\right)\times\rr'.
\end{align}
Now one can define the vector $\OM_1'(t)=\mathbf{\hat R}(t)^{-1}\OM_1(t)$. The application of $\mathbf{\hat R}^{-1}(t)$ on the un-primed vectors, such as in the case of $\rr(t)$, produces the coordinates of the vector in the toggling frame. Therefore $\OM_1'(t)$ is the error vector viewed in the toggling frame. Finally, one obtains the equation of motion for the Bloch vector in the toggling frame,
\begin{align}
    \diff{\rr'(t)}{t}&=\OM_1'(t)\times\rr'(t),
\end{align}
with the time dependence restored in notation.

\section{Convergence of the series $\rr'(t)=\sum_{n=0}^{\infty}\rr'_n(t)$}
\label{sec.convergence}
We obtained the following set of expressions for the terms of the series:
\begin{align}
    \rr_0'(t)&=\rr_0(0)\\
    \rr_1'(t)&=\int_0^t\OM_1'(s)\times\rr_0(0) \mathrm{d}s\\
    \rr_2'(t)&=\int_0^t\OM_1'(s)\times\rr_1'(s) \mathrm{d}s\\
    &...\nonumber
\end{align}
We now show that the series $\rr'(t)=\sum_{n=0}^{\infty}\rr_n'(t)$ is absolutely convergent. Take the norm of each term of the series and establish its upper bound. Trivially, $||\rr_0'(t)||=||\rr_0(0)||$. For the first-order term, the upper bound for the norm is
\begin{align}
	||\rr'_1(t)|| &= \left|\left|\int_0^t\OM_1'(t)\times\rr_0(0)\mathrm{d}s\right|\right| \\
	&\leq\int_0^t||\OM_1'(t)\times\rr_0(0)||\mathrm{d}s \\
	&\leq\int_0^t||\OM_1'||_{\mathrm{max}}||\rr_0(0)||\mathrm{d}s \\
	&=||\rr_0(0)||\ ||\OM_1'||_{\mathrm{max}}t.
\end{align}
For the second-order term, the norm is
\begin{align}
	||\rr'_2(t)|| &= \left|\left|\int_0^t\OM_1'(t)\times\rr'_1(s)\mathrm{d}s\right|\right| \\
	&\leq\int_0^t||\OM_1'(t)\times\rr'_1(s)||\mathrm{d}s \\
	&\leq\int_0^t||\OM_1'||_{\mathrm{max}}||\rr'_1(s)||\mathrm{d}s \\
	&\leq\int_0^t\left(||\OM_1'||_{\mathrm{max}}\right)^2||\rr_0(0)||\ s\ \mathrm{d}s \\
	&=||\rr_0(0)||\ \frac{1}{2}\left(||\OM_1'||_{\mathrm{max}}t\right)^2.
\end{align}
Continuing the calculation, we find that for the $n$th term in the series, the upper bound is
\begin{equation}
	||\rr'_n(t)|| \leq ||\rr_0(0)||\ \frac{1}{n!}\left(||\OM_1'||_{\mathrm{max}}t\right)^n.
\end{equation}
Clearly the power series
	$$\sum_{n=0}^{\infty}||\rr_0(0)||\ \frac{1}{n!}\left(||\OM_1'||_{\mathrm{max}}t\right)^n$$
is convergent for all (finite) $t$. Hence by the comparison test, the series $\sum_{n=0}^{\infty}||\rr'_n(t)||$ also converges. Any absolutely convergent series is also convergent, hence the infinite sum $\rr'(t)=\sum_{n=0}^{\infty}\rr'_n(t)$ converges.

We can use the established upper bound to roughly estimate the error introduced by truncating the series at the $n$th order, which is approximately $||\rr_0||\frac{1}{n!}(||\OM_1'||_{\mathrm{max}}t)^n$. For the truncated series to be a good approximation, we must have 
\begin{align*}
	\frac{1}{n!}(||\OM_1'||_{\mathrm{max}}t)^n &\ll 1 \\
	(||\OM_1'||_{\mathrm{max}}t)^n &\ll n! \approx \left(\frac{n}{e}\right)^n \\
	||\OM_1'||_{\mathrm{max}}t &\ll \frac{n}{e}.
\end{align*}
Since they are different coordinate frame representations of the same vector, $\mathbf{\Omega}_1(t)$ and $\mathbf{\Omega}_1'(t)$ has the same magnitude. Crudely, the first- and second-order approximations are valid for
	$$||\mathbf{\Omega}_1||_{\mathrm{max}}t\ll1.$$

\section{Relationship between $\OM_1'(t)$ and $\pp_i$ for Detuning Error.}
\label{sec.detuning-first-order}
\begin{figure}[t]
    \includegraphics[scale=1]{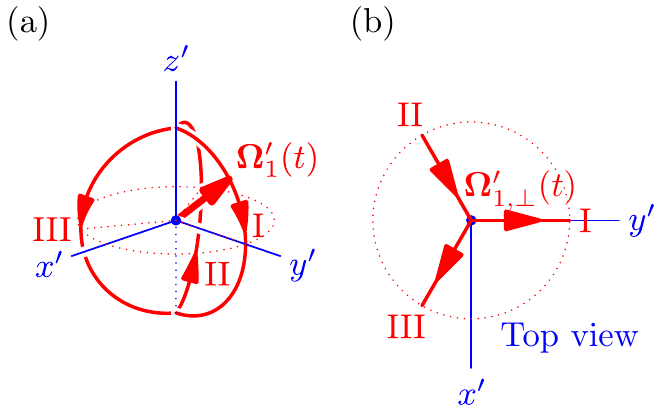}
    \caption{(a): time-dependence of detuning error $\OM_1'(t)$ for the three-step $\pi$-pulse sequence, by symmetry argument. The labels I, II, and III refer to the three steps of the sequence. Note in particular the direction of traversal labelled on each arc, especially for step II. At $t=0$, $\OM_1'$ must point in the $z'$ direction because at this point the toggle frame axes coincide with the lab frame axes.  (b): top view.}
    \label{fig:dOm-detuning-threestep}
\end{figure}

Since we are only considering composite pulses consisting of three $\pi$-pulses, the trajectory of the error $\OM_1'(t)$ for the detuning case must consist of three semi-circular arcs. By symmetry, for the integral in Eq.\eqref{pt-definition} to vanish and the walk to close, one may postulate that the planes of the three arcs must make angles of $2\pi/3$, so that the figure formed has its center of mass located at the origin, as shown in figure \ref{fig:dOm-detuning-threestep}.

\begin{figure}[t]
    \includegraphics[scale=1]{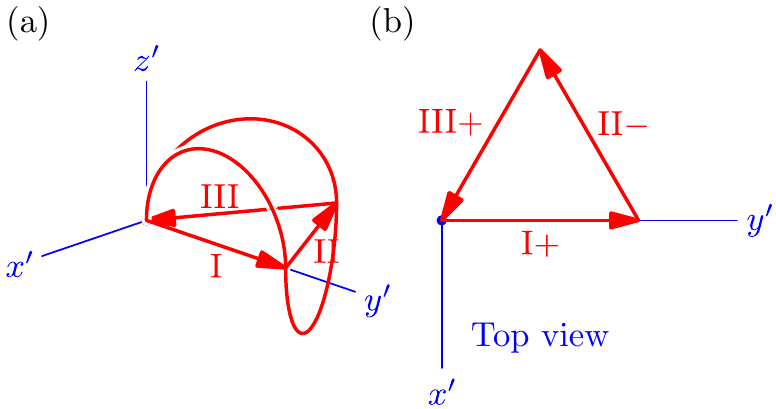}
    \caption{The detuning error integral vectors $\pp_i$ in the three-step sequence $R_0^{\pi}\rightarrow R_{\pi/3}^{\pi}\rightarrow R_0^{\pi}$. (a): each $\pi$-pulse produces a semicircular trajectory whose net effect is a translation along straight lines in the $x'y'$-plane, forming the three error integral vectors. (b): the same curve viewed from the top. A plus or a minus sign is marked depending on whether the curve threads above or under the $z=0$ plane.}
    \label{fig:pt-detuning-threestep}
\end{figure}

It is necessary to draw this previous reasoning for the detuning case back to the equilateral triangle we began with. Treating the $\OM'_1(t)$ as the time-dependent velocity of a phantom particle performing uniform circular motion, it follows from figure \ref{fig:pt-detuning-threestep} that the three error integral vectors so generated indeed form an equilateral triangle, and $\OM'_1(t)$ is the unique way this can happen. 
Notice in particular that the angle formed between the planes of the semi-circles in figure \ref{fig:dOm-detuning-threestep} coincides with the angles formed between $\pp_i$ and the $x'$-axis in figure \ref{fig:pt-detuning-threestep}.

Some additional work needs to be done to read off the pulse direction required in the toggling frame from figure \ref{fig:dOm-detuning-threestep}. Recall that in case of detuning error, the apparent motion of $\OM_1'(t)$ in the toggling frame, such as that in figure \ref{fig:dOm-detuning-threestep} is caused by the rotation of the toggling frame itself, while the vector $\OM_1=\Delta\ \mathbf{\hat{z}}$ in truth stays constant in the lab frame. To generate the apparent rotation of $\OM_1'(t)$ as seen in step I in the toggling frame, the pulse direction must be in the $x'$ direction--rather than in the $-x'$ direction, as one might expect if one naively uses the right-hand-rule on arc I. Similar arguments can be carried out for step II and III. As a result, the pulse directions must form the following angles with the $x'$-axis during the three steps respectively: $0\rightarrow-\pi/3\rightarrow-2\pi/3$.

Now compare these angles with those formed between $\pp_i$ and the $x'$-axis in figure \ref{fig:pt-detuning-threestep}. If one takes the angles from the pulse direction of odd-numbered steps in the toggling frame and add $\pi/2$, one obtains the angle for $\pp_i$. For even-numbered steps (step II only, in this case) addition by $\pi/2$ is replaced by subtraction by $\pi/2$. This relationship will help us generalize to longer sequences. The differentiated treatment between even- and odd-numbered steps can be explained by the fact that the direction of traversal of the semi-circles in figure \ref{fig:dOm-detuning-threestep} toggles between top-down and bottom-up.

\section{Manipulating the Second-order Constraint.}
\label{sec.second-order}
Substitute the definition of $\pp(t)$ into the second-order constraint to obtain
\begin{align}
    \rr_2'|_{t=t_{\mathrm{f}}}&=\int_0^{t_{\mathrm{f}}}\dot\pp(s)\times(\pp(s)\times\mathbf{\hat u})\dd s  \nonumber\\
    &=\int\left[(\dot\pp\cdot\mathbf{\hat u})\pp-(\dot\pp\cdot\pp)\mathbf{\hat u}\right]\dd s,
    \label{second-order-intermediate}
\end{align}
where the simplified notation $\mathbf{\hat u}=\rr_0(0)$ is used for the initial condition. The second term is easily integrated since $\dot\pp\cdot\pp=\frac{1}{2}\diff{}{t}\pp^2$. The first term contains such terms as $p_x\dd p_z$, and $p_y\dd p_z$, which motivates us to consider the area integral $\frac{1}{2}\int\mathbf{p}\times\dot{\mathbf{p}}\ \dd s$. Using vector identities and the fact that the $\pp(t)$ curve is closed, Eq.\eqref{second-order-intermediate} can be manipulated by following the logic below,
\begin{align}
    \mathbf{\hat u}\times(\pp\times\dot\pp)
    &=(\dot\pp\cdot\mathbf{\hat u})\pp-(\pp\cdot\mathbf{\hat u})\dot\pp  \nonumber\\
    &=2(\dot\pp\cdot\mathbf{\hat u})\pp-(\dot\pp\cdot\mathbf{\hat u})\pp-(\pp\cdot\mathbf{\hat u})\dot\pp \nonumber\\
    &=2(\dot\pp\cdot\mathbf{\hat u})\pp-\diff{}{t}\left[(\pp\cdot\mathbf{\hat u})\pp\right]   \nonumber\\
    (\dot\pp\cdot\mathbf{\hat u})\pp&=\mathbf{\hat u}\times\frac{1}{2}(\pp\times\dot\pp)+\frac{1}{2}\diff{}{t}\left[(\pp\cdot\mathbf{\hat u})\pp\right].
\end{align}
Hence, 
\begin{align}
    \rr_2'|_{t=t_{\mathrm{f}}}&=\mathbf{\hat u}\times\frac{1}{2}\int_0^{t_{\mathrm{f}}}\pp\times\dot\pp\ \dd s+\frac{1}{2}\left[(\pp\cdot\mathbf{\hat u})\pp-\pp^2\mathbf{\hat u}\right]_{s=0}^{t_{\mathrm{f}}}  \nonumber\\
    &=\mathbf{\hat u}\times\frac{1}{2}\int_0^{t_{\mathrm{f}}}\pp\times\dot\pp\ \dd s+\frac{1}{2}\left[\pp\times(\pp\times\mathbf{\hat u})\right]_{s=0}^{t_{\mathrm{f}}}.
\end{align}
If the composite pulse already fully compensates for either kind of first-order error, then the $\pp(t)$ curve for said error is necessarily closed, and the boundary term in the integral vanishes. One obtains
\begin{equation}
    \rr_2'|_{t=t_{\mathrm{f}}}=\mathbf{\hat u}\times\left(\frac{1}{2}\oint\pp\times\dd\pp\right).
\end{equation}

\section{Five-step Sequences.}
\label{sec.five-step-sequences}

\begin{figure}[t]
    \includegraphics[scale=1]{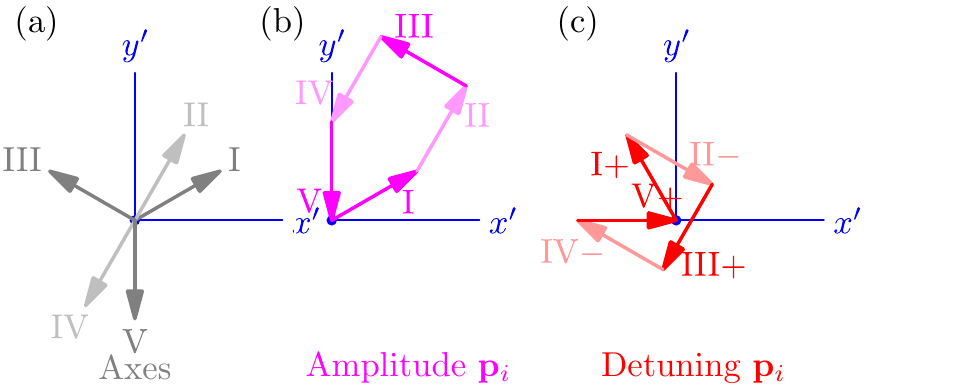}
    \caption{(a) The five pulse directions, (b) amplitude and (c) detuning error integral vectors $\pp_i$ for the Knill sequence in the toggling frame. The $\pp_i$ vectors form five-step closed walks in both cases. The even steps are colored lighter.}
    \label{fig:pt-knill}
\end{figure}

As an illustration of the mapping between the propagator and perturbation approaches, and as another application of the toggling formula Eq.\eqref{toggle-phases-eq}, we consider here the Knill sequence \cite{souza_robust_2011}, the five-step sequence, mentioned in the Introduction, that provides effective amplitude and detuning error suppression. In our notation, the Knill sequence takes the form,
\begin{equation}
    R_{\pi/6}^{\pi}\rightarrow R_{0}^{\pi}\rightarrow R_{\pi/2}^{\pi}\rightarrow R_{0}^{\pi}\rightarrow R_{\pi/6}^{\pi}.
    \label{knill-spec}
\end{equation}
Using Eq.\eqref{toggle-phases-eq} this translates to toggling frame phases $\pi/6\rightarrow \pi/3\rightarrow 5\pi/6\rightarrow -2\pi/3\rightarrow -\pi/2$. The corresponding five $\pp_i$ vectors for amplitude and detuning errors are shown in figure \ref{fig:pt-knill}. The five-step walks are closed in both cases. The Knill sequence is thus fully-compensating to first-order with respect to both amplitude and detuning error. Note that the error integral vectors of the odd steps (i.e., steps I, III, and V) still form an equilateral triangle while the even steps (i.e., steps II and IV) are anti-parallel and thus cancel.

The same algorithm can be used to identify other five-step sequences composed of $\pi$-pulses that share these same properties. Treat the five pulse-direction angles $\phi_i'$ as variables. According to the algorithm, the conditions for full compensation of amplitude and detuning error to first-order are 
\begin{align}
    \mathrm{e}^{\mathrm{i}\phi'_1}+\mathrm{e}^{\mathrm{i}\phi'_2}+\mathrm{e}^{\mathrm{i}\phi'_3}+\mathrm{e}^{\mathrm{i}\phi'_4}+\mathrm{e}^{\mathrm{i}\phi'_5}&=0, \ \\
    \mathrm{i}\mathrm{e}^{\mathrm{i}\phi'_1}-\mathrm{i}\mathrm{e}^{\mathrm{i}\phi'_2}+\mathrm{i}\mathrm{e}^{\mathrm{i}\phi'_3}-\mathrm{i}\mathrm{e}^{\mathrm{i}\phi'_4}+\mathrm{i}\mathrm{e}^{\mathrm{i}\phi'_5}&=0.
\end{align}
The two conditions are obeyed if
\begin{align}
    \mathrm{e}^{\mathrm{i}\phi'_1}+\mathrm{e}^{\mathrm{i}\phi'_3}+\mathrm{e}^{\mathrm{i}\phi'_5}&=0, \\\
    \mathrm{e}^{\mathrm{i}\phi'_2}+\mathrm{e}^{\mathrm{i}\phi'_4}&=0.
\end{align}
The geometric meaning of these conditions is as follows: the three $\pp_i$ vectors for odd $i$ must form an equilateral triangle while the two $\pp_i$ vectors for even $i$ must be anti-parallel. Ignoring inversion and global rotation, one is left with an undetermined relative angle $\alpha$ between the odd and even steps. In the toggling frame, the pulse directions are $\pi/6\rightarrow (\pi/3+\alpha)\rightarrow 5\pi/6\rightarrow (-2\pi/3+\alpha)\rightarrow -\pi/2$, where $\alpha=0$ corresponds to the original Knill sequence. 

\begin{figure}[t]
    \includegraphics[scale=1]{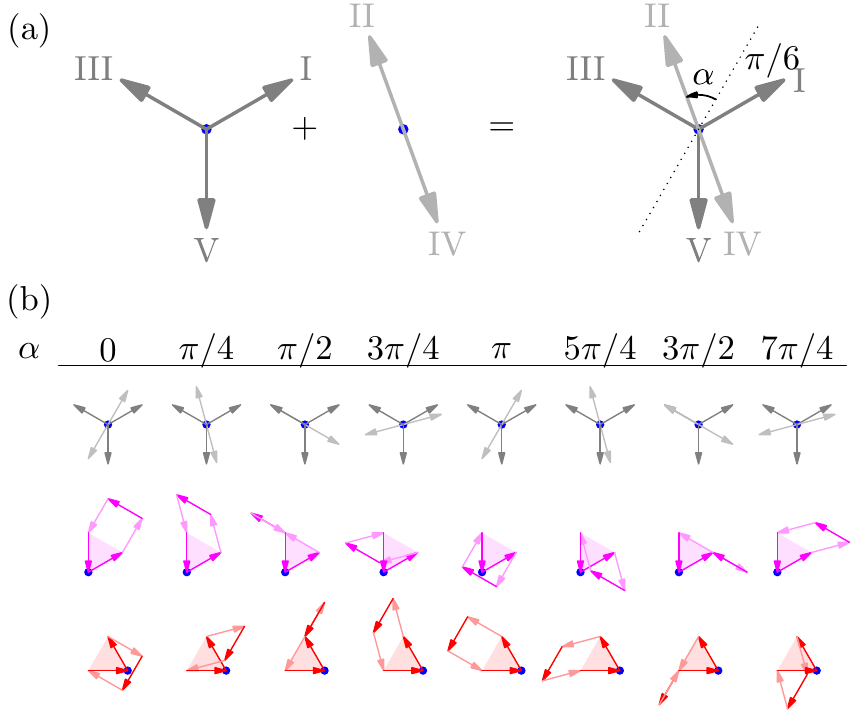}
    \caption{(a): definition of the undetermined angle $\alpha$ of the Knill sequence family, with $\alpha = 0$ the original Knill sequence. (b, top row): examples of the pulse directions, amplitude and detuning $\pp_i$ for different $\alpha$. Sections of the walks that are invariant under changes of $\alpha$ are highlighted. (b, middle row): amplitude error walk. (b, bottom row): detuning error walk. Note the geometrical similarity of amplitude and detuning walks as $\alpha$ varies.}
    \label{fig:knill-freedom}
\end{figure}
The corresponding sequence in the lab frame is
\begin{equation}
    R_{\pi/6+2\alpha}^{\pi}\rightarrow R_{\alpha}^{\pi}\rightarrow R_{\pi/2}^{\pi}\rightarrow R_{-\alpha}^{\pi}\rightarrow R_{\pi/6-2\alpha}^{\pi}.
    \label{knill-spec-general}
\end{equation}
This should be compared with the family of five-step composite pulses that was obtained using the full machinery of propagators (see Eq.(47) of ref.\cite{jones_designing_2013}). The angles that determine the pulse directions are:
\begin{equation}
\bm{\phi}=(\pi+2\alpha',\alpha',-\pi/3,-5\pi/3-\alpha',-7\pi/3-2\alpha') \label{jones-4}
\end{equation}
depending on a single parameter $\alpha'$. All members of this family are fully-compensating to first-order against amplitude and detuning errors. When $\alpha'=7\pi/6$, the Knill sequence results. Although the $\alpha$ parameter of Eq.\eqref{knill-spec-general} is shifted with respect to $\alpha'$, it is clear that the two expressions describe the same family, except that all angles differ by a constant $7\pi/6$.

\begin{figure}[t]
    \includegraphics[scale=1]{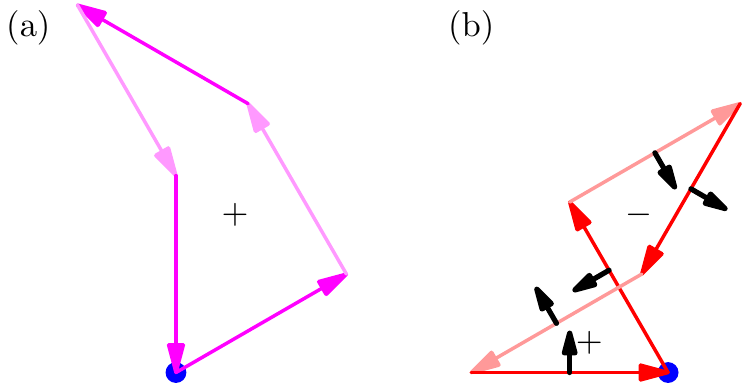}
    \caption{The vector area enclosed by the five-step walk for $\alpha=\pi/3$. (a): top view of the planar five-step walk for amplitude error. The vector area is non-zero and directed along $z'$. (b): top view of the non-planar detuning $\pp(t)$ curve. This polygon also has non-zero area for most $\alpha$. The red polygon in this picture has a near-zero area. Black: the positive direction of the area contribution from the five semi-circles in the detuning $\pp(t)$ curve. Blue: the origin of the toggling frame.}
    \label{fig:pt-ampl-detun-knill-general-second-order}
\end{figure}

As a second illustration, the second-order constraint is applied to the family of five-step sequences that obeys the first-order constraint. The five $\pp_i$ vectors of the five-step closed walks for $\alpha$ equal to $\pi/3$ are shown in figure \ref{fig:pt-ampl-detun-knill-general-second-order} both for amplitude (a) and detuning error (b). 
For the case of amplitude error, the $\pp(t)$ curve is strictly planar so the direction of the vector area is in general along the $z'$ direction. The area of the enclosed pentagonal figure is its magnitude. The second-order constraint for amplitude error is thus not obeyed for $\alpha=\pi/3$. 

\begin{figure}[t]
    \includegraphics[scale=1]{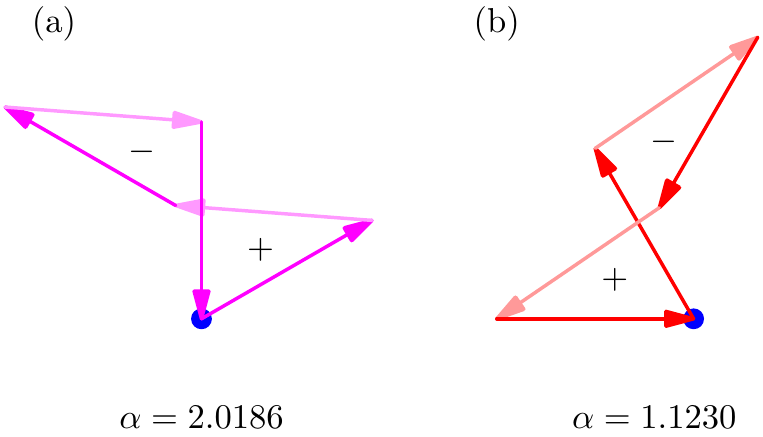}
    \caption{Polygons associated with second-order fully-compensating five-step sequences. Both polygons have zero net area. (a): The sequence with  $\alpha=2.0186$ is fully-compensating against amplitude error. (b): the sequence with $\alpha=1.1230$ is fully-compensating against detuning error. The $+$/$-$ signs indicate the signed contribution to the total vector area of the enclosed figure. Blue dot: the origin of the toggling frame.}
    \label{fig:pt-ampl-detun-knill-uniform-second-order}
\end{figure}

For the case of the detuning error, the family of five-step Knill-like sequences has the property that the vector area contributed by the five semi-circles in the detuning $\pp(t)$ curve exactly cancel (due to the separate cancellation among odd steps and even steps). Only a pentagonal area is left. As shown in figure \ref{fig:pt-ampl-detun-knill-general-second-order}, for $\alpha=\pi/3$ the pentagon is self-intersecting, breaking up the figure in two separate parts. The vector areas of the two parts have opposite signs. For $\alpha=\pi/3$, the two contributions nearly cancel. This suggests that for general $\alpha$ the five-step sequence is not fully-compensating but there could be a special value of $\alpha$ close to $\alpha=\pi/3$ for which there is an exact cancellation. As shown in figure \ref{fig:pt-ampl-detun-knill-uniform-second-order}, this is indeed the case for  $\alpha=\pm\left(\pi-\arccos\left(-\frac{\sqrt{3}}{4}\right)\right)=\pm1.1230$. We will call this a ``magic angle". The value of the magic angle is calculated by equating the area of the parallelogram formed by steps II, III, and, IV of the five-step sequence to the area of the equilateral triangle formed by steps I and V. Similarly, for $\alpha=\pm\arccos\left(-\frac{\sqrt{3}}{4}\right)=\pm2.0186$, the amplitude $\pp(t)$ curve has zero vector area, and the corresponding sequence is fully-compensating to the second-order with respect to amplitude error. These two magic angles agree with the ones reported in \cite{jones_designing_2013}, up to a shift due to the definition of $\alpha$.

\bibliography{References}

\end{document}